\magnification\magstep1
\documentstyle{amsppt}
\input epsf
\hsize=12.5cm
\hoffset=.75cm
\vsize=17cm
\voffset=1.5cm
\parindent=1.5em
\parskip= 2pt plus 4pt
 
\TagsOnRight
\define\ket#1{\vert#1\rangle}
\define\pd#1#2{\frac{\partial#1}{\partial#2}}
\define\pois#1#2{\{#1,#2\}}            
\define\R{{\Bbb R}}                    
\define\BL{{\Bbb L}}                   
\define\w{{\omega}}                    
\define\g{{\gamma }}                   
\define\G{{\Gamma }}                   
                    
\define\f(#1,#2){\frac {#1}{#2}}
\define\dT#1{d_{\bT(#1)}}
\define\bT(#1){{\bold T}^{(#1)}}                                                        
\define\Ker{\operatorname{Ker}}        
\define\J#1{J^{#1}\pi}                 
\define\at#1{{}_{\big\vert_{\ssize #1}}}

\def\refno#1.#2\par{\smallskip   
	  \item{\lbrack#1\rbrack}
		#2\par}
\define\set#1{\{\,#1\,\}}              
\define\<#1>{\langle#1\rangle}         
\define\>#1{{\bold#1}}                 
\define\[{\ifhmode\ \fi$[\mkern-2mu[$} 
\define\]{$]\mkern-2mu]$\ }            
\define\X{{{\goth X}}}                 



\def\fd{\hbox to 30pt{\rightarrowfill}}

\def\izab #1{\Big\downarrow \llap{$\vcenter{\hbox to 50pt{\hfil$\scriptstyle
#1$\ \ }}$}}
\def\izar #1{\Big\uparrow \llap{$\vcenter{\hbox to 50pt{\hfil$\scriptstyle
#1$\ \ }}$}}


\def\supobl #1{\llap{$\vcenter{\hbox to 15pt{$\scriptstyle #1$}}$}}

\def\Saletan{1}
\def\AbM{2}
\def\GS{3}
\def\Bamberg{4}
\def\SW{5} 
\def\Wolfuno{6}
\def\Dragtuno{7}
\def\Dragtdos{8}
\def\Dragttres{9}
\def\DForest{10}
\def\Wolfdos{11}
\def\CL{12}
\def\DFW{13}
\def\Pepin{14}
\def\Pepindos{15}
\def\Marsden{16}
\def\HolmKo{17}
\def\HolmWo{18}
\def\HolmKov{19}
\def\Wolftres{20}
\def\CN{21}
 \def\CP{22}
\def\subrayar#1{$\underline{\smash{\hbox{#1}}}$}
\overfullrule=0 pt
\topmatter

\title
 On the Symplectic structures arising in Geometric Optics 
\endtitle

\author
 Jos\'e F. Cari\~nena$^\dag$ and  
 Javier Nasarre$^\star$  
\endauthor

\affil
 $^\dag$Departamento~de F\'\i sica Te\'orica, Universidad de Zaragoza,\\ 
 50009 Zaragoza {\smc(Spain)} \\
 $^\star$Seminario~de Matem\'aticas, IFP Miguel Catal\'an,
 \\ 
 50009 Zaragoza {\smc(Spain)}
\endaffil

\abstract
Geometric optics is analysed using the techniques of Presymplectic Geometry.
We obtain  the symplectic
structure of the space of light rays in a medium of a non constant refractive 
index by 
reduction from a presymplectic structure,
and using adapted coordinates, we find Darboux coordinates. 
The theory is  illustrated 
 with some examples and we point out  some simple physical applications
\endabstract
\leftheadtext\nofrills{J.F. Cari\~nena and  J. Nasarre}
\rightheadtext\nofrills{Symplectic structures in Optics}
\endtopmatter
\vfil
\hrule
To appear in Forts. der Phys.
\eject
\document
{\baselineskip=.8 cm    
\head
1. Introduction
\endhead

The Hamiltonian formulation of Classical Mechanics is introduced in most 
textbooks as a method of transforming 
the set of second order Euler-Lagrange differential equations 
into a first order system 
(see e.g. \cite{\Saletan}).
 However, systems described by 
Hamilton--like equations are interesting not only in Physics but also  in 
Mathematics, Chemistry, Biology, etc..., and this caused a big interest in 
the study of
these systems. Moreover, the introduction in a geometric way of the concept of
Hamiltonian Dynamical System \cite {\AbM}, which reduces in simple cases to the Hamiltonian 
or the Lagrangian formulation of the Mechanics, has motivated the 
development and  study of symplectic structures and related tools. This
enables us to deal, in a coordinate free way, with mechanical systems for which the configuration space is 
not topologically trivial or even it is an infinite--dimensional linear space 
as it happens in the case of Field Theory.

The Hamiltonian treatment of Geometric Optics, based on Fermat's
principle,
preceded its applications to Classical Mechanics, based on the corresponding 
Hamilton principle. This suggested  that a Hamiltonian formulation of 
Geometric  Optics using the tools of Modern Differential Geometry may
also be useful \cite {\GS}. Then, this geometric formulation of Optics is receiving 
much attention during the last ten years and it is now well established.
For instance, a recent book  on Mathematical Methods in Physics \cite {\Bamberg} includes  a 
Chapter IX on matrix and Hamilton methods in Gaussian optics. 
In particular, the development of Lie methods in Optics \cite {\SW--\DForest} has been very
fruitful and allowed a new perspective in the theory of aberrations,
where the symmetry theory imposes \lq\lq selection rules" for the 
aberration coefficients \cite {\Wolfdos--\DFW}. The main difference of Geometric Optics with particle mechanics is
the existence of refracting surfaces separating regions of a different 
constant refractive index. On the other hand, there is no natural
\lq\lq time parameter" for the ray and it is not the parametrized curve
but the trajectory itself, no matter its parametrization, what is actually
relevant in Geometric Optics.

The usual geometric approach to Symplectic Optics 
is based on the fact that an optical system can be seen as a black--box 
relating constant refractive
index regions and then  incident and outgoing rays are 
characterized by appropriate points   $(q_{\text{in}},p_{\text{in}})$ 
and  $(q_{\text{out}},p_{\text{out}})$ of a  phase space in such a way 
that the
optical device can be considered as a canonical transformation in such
phase space. This reminds us what happens in Quantum scattering theory
where $\ket{\text{in}}$ and $\ket{\text{out}}$ states are related by a unitary 
transformation.

It is very often said in physics textbooks that 
active  and passive viewpoints in transformation theory of physical systems
are equivalent. However, this is only true for the simplest case in which 
the manifold describing the system is $\Bbb R^n$. Actually, coordinates
can only be introduced in a local way while a transformation, from the active
viewpoint,  is a global concept. So, if $M$ is a differentiable manifold,
the configuration space of a system, the phase space, or any other similar
thing else, a transformation of $M$ is a diffeomorphism $F:M\to M$. It induces
a change of charts, the chart  
$(\Cal U, \phi)$ becoming $(\Cal U',\phi')$ with 
$\Cal U'=F(\Cal U)$ and  $\phi'=\phi\circ F^{-1}$. On the contrary,
a change of coordinates is a local  concept and it does not produce any diffeomorphism of the manifold 
$M$.

We will first prove that actually  the space of
light rays in a constant index medium is a symplectic manifold and that
Darboux coordinates for these rays are the $q$ and $p$ usually chosen.
Moreover, it will be shown using the techniques of the geometric approach to 
singular Lagrangian systems
that, under very general assumptions, the 
space of light rays in an arbitrary medium is a symplectic manifold, and
if we restrict ourselves to   optical systems such that
the refractive index depends only on the third coordinate $x^3$ and  
 the index takes, possibly  different, constant values   for $x^3>L$ and
 $x^3<L$, we can choose
Darboux coordinates by fixing the  $x^3$ coordinate  in any of these 
two regions
and taking Darboux 
coordinates for the corresponding problems of constant index. This property
provides a justification of 
the choice of coordinates as usually done for the  ingoing and outgoing 
light rays in the
corresponding constant index media. 
Therefore, it is the choice of \lq\lq Darboux coordinates" in the ingoing
and outgoing regions where light moves freely (i.e., in a constant 
refractive index medium) what motivates that the passage of light through the
optical system seems to be a canonical transformation (active point of view)
when actually it only 
corresponds to a change from one to another Darboux coordinate system
(passive viewpoint) in a symplectic manifold.

The paper is organized as follows: For the sake of completeness we give
in Section 2
a short summary of important concepts of modern Differential Geometry,
which are   well konwn only for few opticists, 
and introduce the notation to be used in the paper.
In Section 3 we develop an idea 
introduced in \cite {\Pepin} to show in a geometric way that the space of 
oriented
straightlines in the  Euclidean three--dimensional space $\Bbb R^3$ is
the tangent bundle $TS^2$ of the two--dimensional sphere, 
and then the identification with
$T^*S^2$, given by the Euclidean metric, shows that this space is endowed
in a natural way with a symplectic structure. We will find out Darboux
coordinates, which essentially reduce to those used by Dragt, Wolf
and coworkers for the description of light rays in a constant  refractive 
index medium. In Section 4 we show how the minimal optical length Fermat's
principle leads to a symplectic formulation of Geometric Optics, by making use 
of the singular optical Lagrangian and its relation with a regular
Lagrangian. The reduction theory  of presymplectic manifolds, following
the ideas   developed by Marsden and Weinstein  \cite{\Marsden},
will be used, and we will illustrate  the method for finding 
coordinates adapted to the distribution defined by the kernel $\Ker \w_L$ 
of the presymplectic structure defined by the singular optical Lagrangian 
in the case of a constant index medium in both cartesian and cylindrical 
coordinates. The most general case of a variable index medium  
is considered in Section 5 and the complete solution is obtained 
for the particular cases of systems in which either the index $n$   
depends on the third coordinate $x^3$ alone  or  the very interesting 
case in which 
the system is
axis--symmetric and $n$
is a function of the distance to this axis \cite {\HolmKo}.
Once Darboux coordinates have been found we can consider 
the problem from the active viewpoint and take advantage of the 
algebraic methods recently developed for computing aberrations.
 
\head
2. Notation and basic definitions
\endhead

The existence  of constraints for mechanical systems motivates that the
configuration space is not $\Bbb R^N$ anymore but some subset $Q$ of points,
called configuration space. 
Physicists  know the convenience of  using 
  generalized coordinates for the description 
of such systems. The corresponding geometric  concept is that  of 
differentiable  
manifold. We recall that a chart  for a topological 
space $M$ is a pair $({\Cal U},\phi)$ where ${\Cal U}$ is an open set
of $M$, $\phi({\Cal U})$ is an open set of $\Bbb R^n$ 
and $\phi:{\Cal U}\to \phi({\Cal U})$ is an homeomorphism. Two charts   
$({\Cal U},\phi)$ and $({\Cal U}',\phi')$ are said to be compatible 
if either 
${\Cal U}\cap {\Cal U}'=\emptyset$ or ${\phi'}\circ \phi^{-1}:
\phi({\Cal U}\cap {\Cal U}')\to\phi'({\Cal U}\cap {\Cal U}')$ is a 
diffeomorphism  
of open sets in $\Bbb R^n$. A differentiable structure in $M$
is given by an atlas, a set of compatible charts covering $M$. From the 
intuitive point of view that means  
that there is a way of parametrizing points by domains in $M$ and coordinates for 
overlapping domains are related by differentiable expressions. The set $M$
endowed with a differentiable structure is said to be a differentiable 
manifold. We will say that a function $f:M\to \Bbb R$  is differentiable 
in a point $m\in M$ when there is a chart $({\Cal U},\phi)$ such
that  $m\in {\Cal U}$ and $f\circ  \phi^{-1}$ is differentiable. In a similar way
a function $F:M\to N$ between differentiable manifolds is 
differentiable at a point  $m\in M$ if there exist charts $({\Cal U},\phi)$   in
$m$ and $({\Cal V},\psi)$ in $F(m)\in N$ such that the map $\psi\circ F\circ \phi^{-1}$
is differentiable. An invertible differentiable map $F$ such that $F^{-1}$
is also differentiable is called diffeomorphism and a diffeomorphism 
$F:M\to M$ corresponds to the physical idea of a transformation of $M$, in 
the so called active point of view. However such a diffeomorphism produces a changes 
of coordinates, the chart 
$(\Cal U, \phi)$ becoming $(\Cal U',\phi')$ with 
$\Cal U'=F(\Cal U)$ and  $\phi'=\phi\circ F^{-1}$. This is called the passive
viewpoint. An expression like $y^i=F^i(x^j)$ can be seen either as a 
transformation 
in which   the point of coordinates $y^i$ is the image under $F$ of the point 
of coordinates $x^i$,  or alternatively as a change of coordinates, from  
$x^i$ to $y^i$, for the same point.

Vectors in a point $m\in M$ may be introduced by an equivalence 
relation among curves 
starting from $m$. So, if $\gamma_1(0)=\gamma_2(0)=m$ are two such curves 
they are equivalent if $\phi\circ\gamma_1$ is tangent to $\phi\circ\gamma_2$
at $\phi(m)$. 
They can alternatively be seen as a map assigning to each differentiable
function in $m$ the real number $\frac d{dt}(f\circ\gamma_1)_{|t=0}$,
and then having in mind that for any curve $\gamma$,
$$\frac d{dt}(f\circ\gamma)_{|t=0}= 
\left(\frac {d(\phi\circ\gamma)^i}{dt}\right)_{|t=0}
\left(\pd{f\circ\phi^{-1}}{x^i}\right)_{|\phi (m)},
$$
the corresponding vector will be denoted $v=v^i\pd{}{x^i}_{|m}$ with $v^i
=\frac {d(\phi\circ\gamma)^i}{dt}_{|t=0}$. Then,   
if $\phi\circ\gamma_1$ is tangent to $\phi\circ\gamma_2$, 
$\frac d{dt}(f\circ\gamma_1)_{|t=0}=\frac d{dt}(f\circ\gamma_2)_{|t=0}$. 
The set of all
vectors in the point $m\in M$ is a linear space called the tangent 
space at $m$ and denoted $T_mM$.

The set of all possible vectors in all points of $M$ is called the tangent
bundle $TM$. It is endowed with a differentiable structure obtained from that
of $M$. In fact, a basis of $T_mM$ is obtained by taking the vectors tangent to 
each coordinate line. The map $\tau:TM\to M$ assigning to each vector 
the point 
where it is placed is such that for any chart in $M$, 
$\tau^{-1}({\Cal U})={\Cal U}\times\Bbb R^n$. The corresponding 
coordinates will be denoted $(q^i,v^i)$. A similar process can be followed
by glueing together the dual spaces  $T^*_mM$ and we will obtain in this way the
so called cotangent bundle $\pi:T^*M\to M$, with coordinates induced from 
base coordinates to be denoted $(q^i,p_i)$. It is remarkable that it is 
possible to use  in 
both spaces  $TM$ and $T^*M$ more general ways of introducing coordinates,
mixturing base and fibre coordinates. However the ones introduced previously
are adapted to the tangent or cotangent structure, respectively, while the more general
coordinates will hide that structure. Sections of these bundles are called vector
fields and 1--forms in $M$, respectively.

The important point is that tangent and cotangent structures are characterized
by the existence of canonical objects. The best known one is a canonical 1--form
$\theta$ in $T^*M$ that in coordinates $(q^i,p_i)$ adapted to the cotangent bundle
looks $ \theta=p_i\,dq^i$. On the other side, the tangent bundle is characterized by
a $(1,1)$--tensor field called vertical endomorphism $S$ that in 
terms of natural 
  coordinates $(q^i,v^i)$ of the tangent bundle $TM$ 
	is given by
$$      S = \pd{}{v^i }\otimes   dq^i.\tag 2.1$$

The exterior differential of the canonical 1--form $\theta$ in a cotangent bundle 
is a closed 2--form
$\omega=-d\theta$ such that $\omega^{\wedge n}\ne 0$, and then this nondegeneracy 
allows us to put in a one to one correspondence vectors and covectors 
in a point, and by extension, vector fields with 1--forms in $M$, in much 
the same way as it happens with Riemannian structures in General Relativity.  In the natural coordinates, 
$\omega=dq^i\wedge dp_i$.
The generalization of this structure is the concept of symplectic manifold, 
a pair $(N,\Omega)$ where $\Omega $ is a nondegenerate closed 2--form 
in the differentiable manifold $N$. The dimension of $N$ is an even number 
$2n$. Moreover, cotangent bundles are 
the local prototype for this more general objects, as  established by
the well known 
Darboux Theorem:
given a point $m\in N$ in a symplectic manifold, there is a coordinate
neighborhood of $m$ with coordinates $(y^1,\ldots,y^{2n})$ 
(called Darboux coordinates)
such that 
$\Omega $ is written $\Omega=dy^i\wedge dy^{n+i}$. Summation from $1$ to $n$ for the index $i$ 
is understood. It is usual to denote these
new coordinates as $y^i=q^i$ and $y^{n+i}=p_i$, for $i=1,\ldots,n$, for 
which the expression looks like for the  symplectic $T^*M$ manifold. 
Of course these Darboux coordinates are not uniquely determined. For instance 
we can introduce new coordinates $P_i=-q^i$, $Q^i=p_i$, that are also 
Darboux coordinates. A change of base coordinates $Q^i=Q^i(q)$ in a cotangent bundle
induces 
a change in fibre coordinates $p_i=P_j\partial Q^j/\partial q^i$ in such a way that
$(Q^i,P_i)$ are also Darboux coordinates.
Notice that  in the general case of an arbitrary symplectic manifold 
the notation of $q$'s and $p$'s is arbitrary while in the
$T^*M$ case $q$'s are actually coordinates on the base space and $p$'s are
fibre coordinates and therefore these last ones should 
take all  possible real values.
Now, the choice of an arbitrary function $H\in C^\infty (N)$, determines
the vector field $X_H$ such that $i(X_H)\omega=dH$, i.e., $\omega(X_H,Y)=YH$, 
$\forall Y\in \X(N)$, that in 
Darboux coordinates   is $$X_H=\pd H{p_i}\pd{}{q^i}-\pd H{q^i}\pd{}{p_i}.
\tag 2.2$$
Its integral curves will be determined in Darboux coordinates 
by the well--known Hamilton equations.

There is no natural symplectic structure in $TM$, but for any  
function $L\in C^\infty(TM)$, we may define an exact 2--form in $TM$,
$\w_L=-d\theta_L  $, with the 1--form $\theta_L$ being defined by 
$\theta_L  =dL\circ S$, and a function $E_L=
\Delta (L)-L$, called energy function. In the above mentioned coordinates of 
$\,TM$ we have the following expressions:
$$      \theta_L   = \pd{L}{v^i }\,dq^i,  \tag 2.2$$
$$      \w_L = \pd{^2L}{q^i\partial v^j} dq^j\wedge dq^i + 
\pd{^2L}{v^i\partial v^j}dq^i\wedge dv^j,       \tag 2.3$$
$$\Delta= v^i \pd{}{v^i }\tag 2.4$$
$$E_L = v^i \pd{L}{v^i } - L,   \tag 2.5$$
Here $\Delta \in \X(TM)$ denotes the Liouville vector field generating
dilations along the fibres. When $\omega_L$ is nondegenerate it defines a  
symplectic structure on $TM$, and a vector field $X_L$  determined by 
$i(X_L)\omega_L=dE_L$. The Legendre transformation ${\Cal F}L:TM\to T^*M$
relates the symplectic structures in $TM$ and $T^*M$ respectively, and 
if $H\in C^\infty(T^*M)$  is defined by $H\circ {\Cal F}L=E_L$, then 
${\Cal F}L_*X_L=X_H$. 

Presymplectic  structures may arise either when using some constants of motion
for reducing the phase space or also when the Lagrangian that has been  
chosen is singular. Then we will have a pair  $(N_0,\Omega_0)$ where $\Omega_0$
is a closed but degenerate 2--form. A consistent solution of the dynamical equation
can only be found in some points, leading in this way to the  final 
constraint submanifold $N$
 introduced by Dirac (see e.g. \cite{\Pepindos}). 
 The pull back  $\Omega$ of the form $\Omega_0$ on this manifold will 
 be assumed to be of constant rank.
The recipe for dealing with these systems was given by Marsden and Weinstein
\cite{\Marsden}.
First, in every point $m\in N$, $\Ker \Omega_m$ is a $k$--dimensional linear 
space, so defining what is called a $k$--dimensional distribution. The important
point is that closedness of $\Omega$ is enough to warrant that the distribution 
is integrable (and then it is called foliation):  for any 
point $m\in N$, there is a $k$--dimensional submanifold of $N$ passing 
through $m$
and such that the tangent space at any point $m'$ of this surface 
coincides with $\Ker \Omega_{m'}$. Such integral $k$--dimensional submanifolds
give a foliation of $N$ by disjoint leaves and in the  case in which the quotient
space $\widetilde N=N/\Ker \Omega$ is a differentiable manifold, then it 
is possible to define a nondegenerate closed 2--form  $\widetilde \Omega $
in $\widetilde N$ such that $\widetilde \pi^*\widetilde \Omega=\Omega$.
Here $\widetilde \pi:N\to \widetilde N$ is the natural projection. 
It suffices to define $\widetilde \Omega(\widetilde v_1, \widetilde v_2)=
\Omega(v_1,v_2)$, where $v_1$ and $v_2$ are tangent vectors to $N$ 
projecting under $\widetilde \pi$ onto 
$\widetilde v_1$ and $\widetilde v_2$ respectively. The symplectic
space $(\widetilde N,\widetilde\Omega)$ is said to be the reduced space.
 
 \head
3. The symplectic structure of the light rays space
 in a constant index medium
\endhead

In a recent paper \cite{\Pepin} it was shown that the set
 of oriented
geodesics of a Riemannian manifold can be endowed with
 a symplectic structure. We will study here some particular examples of interest
 in Symplectic Optics.
To start with, let us see that  the set of 
oriented straightlines in the plane, that is well known to be the set
of light rays  in a two--dimensional constant rank medium, can be endowed with a 
symplectic structure. Moreover, such a set can be considered as the 
cotangent bundle of the one--dimensional sphere $S^1$, i.e. a circumference. 
If an origin $O$
has been chosen in the plane,
 every oriented straightline  that does not pass through the point $O$ 
is characterized by  a unit vector $\bold s$ pointing in the line 
direction and a
 vector $\bold v$ orthogonal to  $\bold s$  with end on the line and
 origin in $O$. 
  \epsfbox{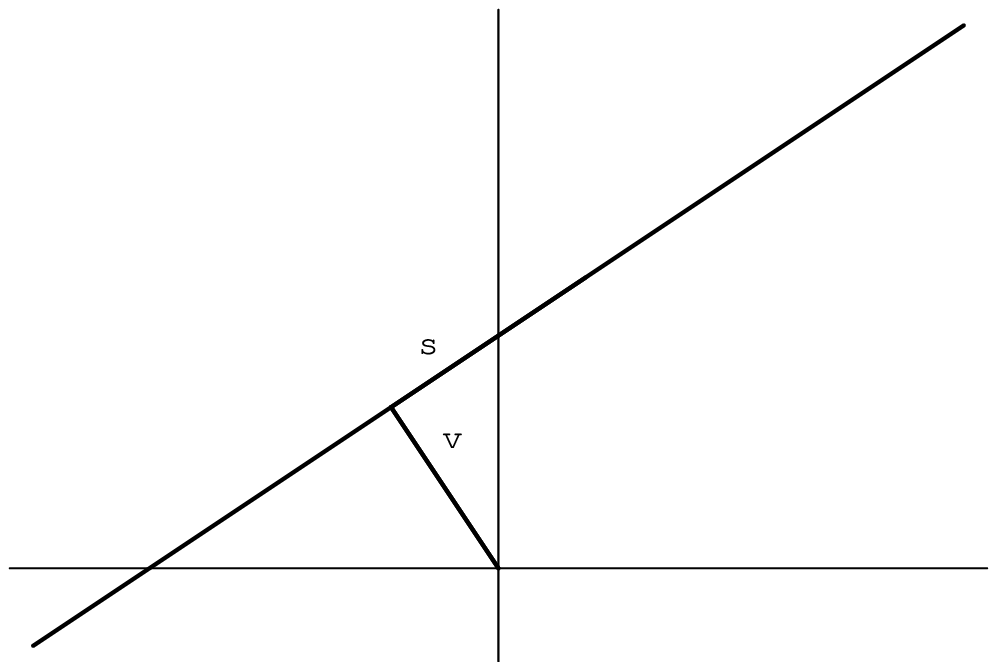}
  \centerline{\boxed{\text{Fig.1}}}

 So, the straightlines of a pencil of oriented parallel lines 
 have the same direction and only differ in the distance to the origin,  
 and therefore they are characterized by proportional vectors $\bold v$ and 
 the same $\bold s$.
Straightlines passing through $O$ with direction given by
 $\bold s$
correspond to $\bold v=0$. The vectors $\bold v$ and
$\bold s$ being orthogonal and $\bold s.\bold s=1$, the couple
$(\bold s,\bold v)$ can be seen as a tangent vector to the 
unit circle $S^1$ at the point described by $\bold s$
as indicated in the following figure:
  
  \epsfbox{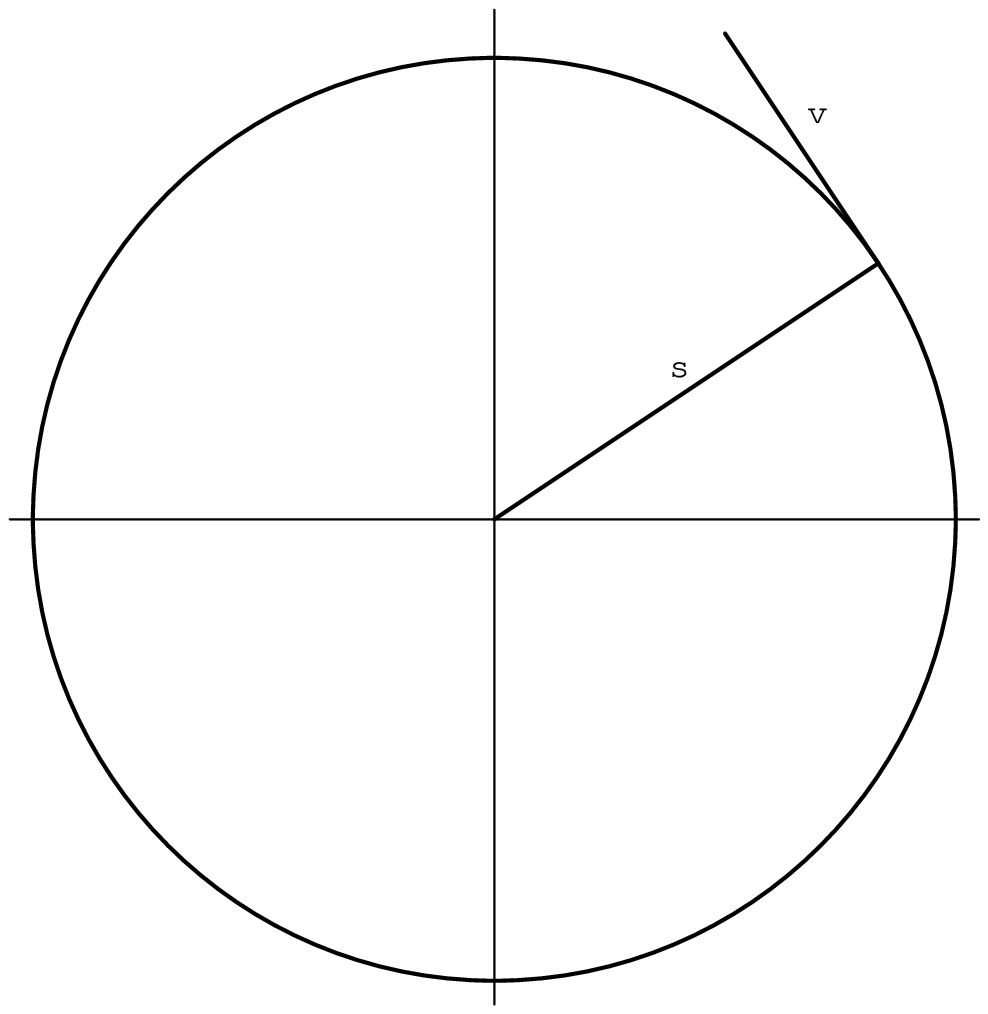}
\centerline{ \boxed{\text{Fig.2}}} 
\medskip

The circle $S^1$ is a Riemannian one--dimensional 
submanifold 
of  the Euclidean two--dimensional space and the
 Riemannian metric in $S^1$ can be used to identify in each point $\bold s$
the tangent space $T_{\bold s}S^1$ with its dual space
$T^*_{\bold s}S^1$ and therefore   the tangent bundle $TS^1$ 
with the cotangent bundle $T^*S^1$. This identification shows us  
that the space of oriented straightlines in the Euclidean two--dimensional 
space can be  endowed with an exact symplectic structure which corresponds 
to the canonical   structure for the cotangent bundle $T^*S^1$. 

Let us look for the kind of Darboux 
coordinates for  such a symplectic form
mentioned in the preceding section. They will be induced from a 
choice of coordinates in the base space. As it is a circle, a good 
choice will be an angle coordinate. 
A straightline 
$y=m\, x+ b$  with slope $m= \tan \theta$ will be represented
by a vector orthogonal to the vector $\bold s=(\cos \theta, 
\sin\theta)$, and length $b\, \cos\theta$,   namely, 
$$\bold v= b \cos \theta\, \pd {}{\theta}.\tag 3.1$$

On the other hand, the vector $\partial/\partial\theta$ is unitary 
in the Euclidean metric, and therefore the point 
$(\theta,p_\theta)\in T^*S^1$
corresponding to   $(\theta, v_\theta)$ is given by $p_\theta=v_\theta$.
Therefore, the symplectic form in $TS^1$ translated from the canonical 
symplectic structure in $T^*S^1$ $\omega_0=d\theta\wedge dp_\theta$ will be 
$$\w= d\theta\wedge d(b\, cos \theta)=d(\sin \theta)\wedge db,\tag 3.2$$
which shows that Darboux coordinates for $\w$ adapted to the cotangent 
structure are not only $(\theta, b\cos\theta)$ but also  
$$q= \sin \theta, \qquad p=b,\tag 3.3$$
which are more appropriate  from the experimental viewpoint. So, the flat screens 
arise here as a good choice for Darboux coordinates.

The choice of the norm  equal to one for the vector $\bold s$  
representative of the line
direction is arbitrary. In the case of Geometric Optics, 
the light rays in a medium of a constant index $n$ are straightlines. For reasons
which will become clear later, the choice usually done 
is $\bold s.\bold s=n^2$, the Darboux coordinate $q$ then being $q=n\, \sin
\theta$. This leads to the image of the Descartes sphere, a sphere of radius
 $n$ whose  points describe the ray directions (see e.g. \cite {\Wolftres}). 
 Therefore, the space of light rays in a medium of constant 
index is like a phase space of a system with configuration
 space the sphere
$S^1$ of radius $n$ corresponding to ray directions. It is noteworthy
that it is very often  used a somehow misleading   notation of $p$  for
 the product $n\sin \theta$ even if  the last one  corresponds to a 
coordinate
in the configuration space.  In our identification it is clear that 
the coordinate $q$ is bounded while $p$ is unbounded, as it was expected 
to be for a cotangent bundle structure.

The study of oriented straightlines in Euclidean 
three--dimensional space follows a 
similar pattern, but  the search for
Darboux coordinates is a bit  more involved. A 
straightline $r$
will
be characterized by a vector  $\bold s$, running a two--dimensional sphere,
 which fixes the ray direction, and a vector  $\bold v$ 
orthogonal to $r$
with origin in $O$ and end in $r$. The conditions 
$\bold s.\bold s=1$ and $\bold s.\bold v=0$ mean that the couple
$(\bold s,\bold v)$ determines an element of the tangent 
bundle $TS^2$.
The points $(\bold s,\bold v)$ of the zero section correspond to 
straightlines passing through 
the origin.
 
The Riemann structure in $S^2$ coming from the Euclidean
 structure in $\R^3$ provides  us 
 the identification of $TS^2$ with $T^*S^2$ and then it allows to endow
 in this way 
$TS^2$, and therefore the set of oriented straightlines in
 $\R^3$, with a symplectic structure coming from the canonical structure in 
$T^*S^2$. In order to find  Darboux coordinates for it  let us consider, for instance,  the local chart in $S^2$
with domain in the upper hemisphere and the map $\varphi (x,y,z)=(x,y)$.
In other words, the point in $S^2$ with local coordinates 
$(u_1,u_2)$ is $(u_1,u_2, \sqrt{1-u_1^2-u_2^2})$. 
The corresponding basis for the tangent space in a point of coordinates 
$(u_1,u_2)$ is 
$$\bold  e_{u_1}=(1,0, -\frac{ u_1}{\sqrt{1-u_1^2-u_2^2}}),\qquad 
\bold e_{u_2}=(0,1, -\frac {u_2}{\sqrt{1-u_1^2-u_2^2}})
$$
and the Riemannian metric will be given in these coordinates by 
$$
g(u_1,u_2)=\frac {1}{1-u_1^2-u_2^2}\pmatrix {1-u_2^2}& 
 {u_1u_2} \\
 {u_1u_2}&{1-u_1^2}
\endpmatrix.\tag 3.4
$$

Then, a vector $\bold v$ tangent to the sphere $S^2$ can be 
expressed either as $\bold v= v_x \bold e_x+v_y \bold e_y+
v_z \bold e_z$ or as $\bold v= a\, \bold  e_{u_1}+b\,\bold  e_{u_2}$.

In particular, it is easy to check that $a=v_x$ and
 $b=v_y$. In the identification of $T_{\bold  s}S^2$ with
$T^*_{\bold  s}S^2$ given by the Riemannian metric, the corresponding
 point will have coordinates
 $$\aligned
p_1&=\frac {1-u_2^2}{1-u_1^2-u_2^2} v_x+
 \frac {u_1u_2}{1-u_1^2-u_2^2}v_y\\p_2&=\frac {u_1u_2}
{1-u_1^2-u_2^2}v_x+
\frac {1-u_1^2}{1-u_1^2-u_2^2} v_y,
\endaligned
$$ 
and therefore the symplectic 2--form in $TS^2$ will be given by
$$
\aligned 
\w&= du_1\wedge d\left(\frac{(1-u_2^2) v_x+u_1u_2v_y}
{1-u_1^2-u_2^2}\right)+du_2\wedge d\left(\frac{u_1u_2v_x+
(1-u_1^2)v_y}{1-u_1^2-u_2^2}\right)\\&=
du_1\wedge d\left(\frac{(1-u_2^2) v_x-u_1^2v_x+u_1^2v_x
+u_1u_2v_y}
{1-u_1^2-u_2^2}\right) \\&+du_2\wedge d\left(\frac{u_1u_2v_x+
(1-u_1^2)v_y-u_2^2v_y+u_2^2v_y}{1-u_1^2-u_2^2}\right)
\\
&=du_1\wedge d\left( v_x-\frac{u_1}{\sqrt{1-u_1^2-u_2^2}}\,
\frac {-u_1v_x-u_2v_y}{\sqrt{1-u_1^2-u_2^2}}\right)
\\&+du_2\wedge
d\left(v_y-\frac{u_2}{\sqrt{1-u_1^2-u_2^2}}\,
 \frac {-u_1v_x-u_2v_y}{\sqrt{1-u_1^2-u_2^2}}\right)
\\
&=du_1\wedge d\left( v_x-\frac{u_1}{\sqrt{1-u_1^2-u_2^2}}
v_z\right)+
du_2\wedge \left(v_y-\frac{u_2}{\sqrt{1-u_1^2-u_2^2}}
v_z\right),
\endaligned\tag 3.5
$$
that can be rewritten as 
$\w= du_1\wedge db_x+du_2\wedge db_y,$
with 
$$
b_x=v_x-\frac{u_1}{\sqrt{1-u_1^2-u_2^2}}v_z\qquad b_y=v_y
-\frac{u_2}{\sqrt{1-u_1^2-u_2^2}}v_z. \tag 3.6
$$

The geometric interpretation is clear. The point of
 cartesian coordinates 
$(b_x,b_y,0)$ is the intersection point of the straightline $r$ with the 
plane $\pi$ defined by $z=0$. In fact the points of   
 $r$ are given by $\bold v+ \lambda \bold s$
for any real number $\lambda\in \R$. The intersection point of $r$ with   $\pi$ 
corresponds
 to the value $\lambda=-\dfrac {v_z}{s_z}$, and then its  coordinates turn out 
to be $(b_x,b_y,0)$. So, in the study of the set of light rays
in a three--dimensional constant index medium, Darboux 
coordinates are,  for instance, given by the projections $\bold e_x.
\bold s$ and  $\bold e_y.\bold s$,  as  $q$--coordinates
 and the cartesian 
coordinates of the intersection point of $r$ with $\pi$ as
momentum coordinates.   Therefore, it also shows that   flat screen are 
appropriate as a  method of introducing 
Darboux coordinates for the symplectic form.
 
\head
4. From Fermat's principle to symplectic geometry
\endhead
Light rays  trajectories in Geometric Optics are determined by Fermat's
principle,
according to which the ray path connecting two points is 
the one making stationary the optical length
$$\delta \int_\g  n\, ds=0.\tag 4.1$$
 This corresponds to 
the well--known Hamilton's principle of Classical Mechanics with an \lq\lq optical 
Lagrangian" $L=n\, \sqrt{v_x^2+v_y^2+v_z^2}$, which is a differentiable 
function in $T\Bbb R^3$ up to at the zero section. 

 The point to be remarked here is that such Lagrangian is a
homogeneous function of degree one in the velocities and consequently $L$
 is singular and the corresponding energy function vanishes identically. The singularity 
of $L$ is related with time reparametrizacion invariance 
(see e.g. \cite {\CN}
for a geometric approach to the second Noether's Theorem). Moreover, it is possible 
to relate the solutions of the Euler--Lagrange equations for $L$ with those of
the regular Lagrangian $\BL=\frac 12 L^2$, up to a reparametrization. The geometric
explanation of this fact was given in \cite {\Pepin}. We will study here the 
presymplectic structure defined by $\w_L$ and the reduction procedure giving rise
to a symplectic structure in the quotient manifold. We will also find Darboux 
coordinates for the reduced symplectic form. For the sake of simplicity, let us
 first consider in this section  the simplest case in which $n$ is constant.

The coordinate expressions of the 1--form $\theta_L$ and the 2--form $\w_L$ are 
(see 
\cite {\Pepin})
$$\aligned
\theta_L&=\frac {n^2}{L}(v_x\, dx+v_y\, dy+v_z\, dz)\\
\w_L&=\frac {n^2}{L}(dx\wedge dv_x+dy\wedge dv_y+dz\wedge dv_z)\\&+
\frac {n^4}{L^3}(v_x\, dv_x+v_y\, dv_y+v_z\, dv_z)\wedge(v_x\, dx+v_y\, dy+v_z\, dz),
\endaligned\tag 4.2
$$
the kernel of $\w_L$ being generated by the Liouville vector field
$$\Delta =v_x\,\pd{}{v_x}+v_y\,\pd{}{v_y}+v_z\,\pd{}{v_z} \tag 4.3$$
and the dynamical vector field giving the dynamics corresponding to $\BL$,
$$
\G= v_x\,\pd{}{x}+v_y\,\pd{}{y}+v_z\,\pd{}{z}.\tag 4.4
$$

 The 2--form $\w_L$ is closed and defines an integrable two--dimensional
distribution $\Cal D$ generated by $\Delta $ and $\Gamma$. 
The relation  $[\Delta, \G]=\G$, expressing that 
the vector field $\G$ is homogeneous of degree one in velocities,
 shows explicitly that $\Cal D$ is an involutive,  and therefore integrable,
 distribution. 

In order to find coordinates adapted to $\Cal D$ we can consider that, in a 
neighbourhood of a point for which $v_z\not =0$,   the distribution  
$\Cal D$ is generated by the commuting vector fields $\Delta$ and 
$K=\dfrac1{v_z}\,\G$. The general
theory of integrable distributions, see e.g \cite{\CP}, says us that there exist local coordinates
$(x_1,x_2,x_3,y_1, y_2, y_3)$ 
such that
$$
\Delta =\pd {}{x_3}\qquad K=\pd {}{y_3}.\tag 4.5
$$

The coordinates $x_1,x_2, y_1$ and $ y_2$ will be given by functions
solution of the system
$$
\left\{ \aligned \Delta f&=v_x\,\pd{f}{v_x}+v_y\,\pd{f}{v_y}+v_z\,\pd{f}{v_z}= 0\\
Kf&=\frac1{v_z}\left( v_x\,\pd{f}{x}+v_y\,\pd{f}{y}+v_z\,\pd{f}{z}\right)=0, 
\endaligned\right .
$$
while $x_3$ will be given by a solution of 
$$\left\{ \aligned \Delta f&=v_x\,\pd{f}{v_x}+v_y\,\pd{f}{v_y}+v_z\,\pd{f}{v_z}= 1\\
Kf&=\frac1{v_z}\left( v_x\,\pd{f}{x}+v_y\,\pd{f}{y}+v_z\,\pd{f}{z}\right)=0,
\endaligned\right.
$$
and $y_3$ by a solution of 
$$\left\{ \aligned \Delta f&=v_x\,\pd{f}{v_x}+v_y\,\pd{f}{v_y}+v_z\,\pd{f}{v_z}= 0\\
Kf&=\frac1{v_z}\left( v_x\,\pd{f}{x}+v_y\,\pd{f}{y}+v_z\,\pd{f}{z}\right)=1. 
\endaligned\right.
$$

A particular solution of these systems is given by 
$$\aligned x_1&=\frac {v_x}{v_z}z-x, \quad x_2=\frac {v_x}{v_z}y-
\frac {v_y}{v_z}x,\quad 
x_3=\log v_z\\
y_1&=\frac {v_x}{v_z},\quad y_2=\frac {v_y}{v_z},\quad y_3=z
\endaligned.\tag 4.6
$$

 The inverse transformation of coordinates is 
$$
\aligned
x&=y_1y_3-x_1, \quad y=\frac {x_2-y_2x_1}{y_1}, \quad z=y_3\\
v_x&=y_1e^{x_3}, \qquad v_y=y_2e^{x_3}, \quad v_z=e^{x_3}
\endaligned.\tag 4.7
$$

When written in these new coordinates the expression of the 2--form $\w_L$
becomes 
$$\aligned \w_L&= \frac {ne^{3x_3}}{L^3}\left[ dy_1\wedge dx_1
 +\frac {1+y_1^2}{y_1} dx_2
\wedge dy_2
+\frac {y_2}{y_1} dy_2\wedge dx_1\right.
\\ &\left.+\frac {x_2-x_1y_2+x_2y_1^2}
{y_1^2}dy_2\wedge dy_1+y_2 dy_1\wedge dx_2\right],
\endaligned 
$$
which can also be rewritten with a reordering of terms as
$$\aligned \w_L&=\frac {n}{(1+y_1^2+y_2^2)^{3/2}} \left[d(-x_1)\wedge ((1+y_2^2)dy_1-
y_1y_2dy_2)\right.\\
&\left.+ d \left(\frac{x_2-y_2x_1}{y_1}\right)\wedge ((1+y_1^2)dy_2-y_1
y_2dy_1)\right],
\endaligned
$$
or 
$$
\w_L= n\, d(-x_1)\wedge d\left(\frac {y_1}{\sqrt{1+y_1^2+y^2_2}}\right)+
n\, d\left(\frac{x_2-y_2x_1}{y_1}\right)\wedge d
\left(\frac{y_2}{\sqrt{1+y_1^2+y^2_2}}\right),\tag 4.8
$$
and when using the old coordinates, 
$$
\aligned 
&-x_1=x-\frac {v_x}{v_z} z=b_x, \qquad 
\frac{x_2-y_2x_1}{y_1}=y-\frac {v_x}{v_z}z=b_y\\
&\frac {n\,y_1}{\sqrt{1+y_1^2+y^2_2}}=\frac {n\,v_x}{\sqrt{v_x^2+v_y^2+v_z^2}}
=p_x\\
&\frac{n\,y_2}{\sqrt{1+y_1^2+y^2_2}}=\frac {n\,v_y}{\sqrt{v_x^2+v_y^2+v_z^2}}=p_y
\endaligned \tag 4.9
$$
from which we have got Darboux coordinates for the reduced form. Notice that
 the 
explicit expression of $\w_L$ in these coordinates shows that it passes to
the 
quotient, because its coordinates do not depend
on $x_3$ and $y_3$, and then there exists a symplectic form $\Omega$ in
the 
quotient space with the same coordinate expresion as in (4.8) for 
$\w_L$. The quotient space is but the space of light rays and therefore it 
is $TS^2$ as indicated before.

We can also use cylindrical coordinates $(r,\theta,z)$ for solving the problem.
The vector fields $\G_{\Bbb L}$ and $\Delta$ are then given by
$$\aligned
\G_{\Bbb L}&= v_r \pd {}r+ v_\theta \pd{}{\theta}+ v_z \pd {}z+  
r v_\theta ^2\pd{}{ v_r} -2\frac{ v_r v_\theta}r \pd {}{ v_\theta}\\
\Delta&=v_r \pd {}{ v_r}+ v_\theta \pd{}{ v_\theta}+ v_z \pd {}{ v_z}
\endaligned \tag 4.10
$$
and a similar computation, choosing now $K'=\frac 1{v_r}\G_{\Bbb L}$
and $\Delta$ as generators of $\Cal D$, leads to the following adapted
coordinates
$$\aligned {x'}_1&=\theta-\arctan \frac {v_r}{rv_\theta},
 \quad {x'}_2=z-\frac {rv_rv_z}{v_r^2+r^2v_\theta^2} , \quad {x'}_3=\log {v_z}\\
 {y'}_1&=r^2\frac{v_\theta}{v_z}, \quad {y'}_2=\frac {v_r^2+r^2v_\theta^2}{v_z^2} , 
\quad {y'}_3=r,
\endaligned
$$
and Darboux coordinates for the induced form in the quotient are
$$\aligned
\xi_1&=\theta-\arctan \frac {v_r}{rv_\theta},\quad  \xi_2=z-\frac {rv_rv_z}{v_r^2+r^2v_\theta^2} \\
\eta_1&=\frac  {nr^2v_\theta}{\sqrt{v_r^2+r^2v^2_\theta+v_z^2}},\quad 
\eta_2=\frac  {n v_z}{\sqrt{v_r^2+r^2v^2_\theta+v_z^2}}.
\endaligned
\tag 4.11
$$
\bigskip
\head
5. The case of a medium of a non--constant index
\endhead
 Let us now consider the most general case in which the refractive index of
the medium is not constant but it is given by a smooth function $n(x^1,x^2,
x^3)$.
Fermat's principle suggests us to consider the corresponding mechanical problem 
described by a singular Lagrangian $L(q,v)=[g(v,v)]^{1/2}$, where $g$ is
 a metric conformal to the Euclidean metric $g_0$,
$$g(v,w)= n^2 g_0(v,w).\tag 5.1
$$

This problem was analysed in \cite{\Pepin} where it was shown that 
its study can be reduced to that of a regular Lagrangian 
$\Bbb L=\frac 12 L^2$. This
Lagrangian $\Bbb L$ is quadratic in velocities and the dynamical 
vector field $\G_{\Bbb L}$ solution of the dynamical equation
$i(\G_{\Bbb L})\w_{\Bbb L}=
dE_{\Bbb L}=d{\Bbb L}$ is not only a second order differential equation vector field
 but, moreover, it is a spray  \cite{\CP}, 
 the projection onto
 $\Bbb R^3$ of its integral curves being the geodesics of the Levi--Civita 
connection defined by $g$. In other words, $\G_{\Bbb L}$ is
the geodesic spray given by
$$\G_{\Bbb L}=v^i\pd {}{q^i}-\G^i\,_{jk}v^jv^k\pd {}{v^i} ,\tag 5.2
$$
where the Christoffel symbols $\G^i\,_{jk}$ are
$$\G^i\,_{jk}=\frac 12 g^{il}\left[ \pd{g_{kl}}{x^j}+\pd{g_{jl}}{x^k}
-\pd{g_{jk}}{x^l}
\right]
$$ with $g^{ij}$ being the inverse matrix of $g_{ij}$.

In the particular case we are considering where $g(v,w)= n^2 g_0(v,w)$,
 and using cartesian coordinates,
$$
\G^i\,_{jk}=\frac 1{n}\left[ \pd n{x^j}\delta^i_k+\pd n {x^k}\delta^i_j-
\pd n{x^i}\delta_{jk}
\right].\tag 5.3
$$

It was also shown in \cite{\Pepin} that the kernel of $\w_L$ is two--dimensional 
and it is generated by  $\G_{\Bbb L}$  and the Liouville vector field $\Delta$.
The distribution $\Cal D$ defined by $\Ker \w_L$ is integrable because
$\w_L$ is closed;
actually $[\Delta,  \G_{\Bbb L}]= \G_{\Bbb L}$ and the distribution is also 
generated by $\Delta$ and 
$K$ defined by $K=\frac 1{v^3}  \G_{\Bbb L}$, for which $[\Delta, K]=0$. In cartesian
coordinates $K$ is expressed as follows:
$$K=\frac 1{v^3}\left[v^i\pd{}{x^i}- \left(\frac 2n v^i( v.\nabla n)-
 \frac{\| v \|^2}{n}\pd n{x^i}\right)\pd {}{v^i}\right].\tag 5.4
$$

 The theory of distributions suggests us the introduction of new local
 coordinates
$y^i=F^i(x,v)$, $i=1,\ldots ,6$, adapted to the distribution $\Cal D$ 
defined
by  $\Ker \w_L$,
i.e., such that $K=\pd {}{y^3}$, $\Delta =\pd {}{y^6}$. The search for 
these new
coordinates is based on the solution of
the partial differential equation system
$$KF^1=1, \quad \Delta F^1=0,\quad KF^2=0, \quad \Delta F^2=1,$$
and 
$$ 
KF^{2+a}=0, \quad \Delta F^{2+a}=0, \, \text{for } a=1\ldots, 4.
$$

The explicit computation of these functions depends very much on
the choice of the function $n(x^1,x^2,x^3)$. We will illustrate next
the theory with some particular examples.

If $n$ only depends on $x^3$, the dynamical vector field is
$$\aligned 
\G_{\Bbb L}&=v^i\pd {}{x^i}-\frac 2n v^1v^3\frac {dn}{dx^3}\pd{}{v^1}-
\frac 2n v^2v^3\frac {dn}{dx^3}\pd{}{v^2}\\&+\frac 1n ({v^1}^2+{v^2}^2-{v^3}^2)
\frac {dn}{dx^3}\pd{}{v^3},
\endaligned
\tag 5.5
$$
and a function $F^1$ solution of $\Delta F^1=0$ should be
$F^1=f^1(x^1,x^2,x^3,u^1, u^2)$ with  $u^1=\frac {v^1}{v^3}$ y 
$u^2= \frac {v^2}{v^3}$, and then 
the condition $KF^1=1$ reads
$$\aligned KF^1&=u^1\pd {f^1}{x^1}+ {u^2}\pd {f^1}{x^2}+
\pd {f^1}{x^3}
-\frac 2n u^1\frac {dn}{dx^3}\pd{f^1}{u^1}-
\frac 2n u^2 \frac {dn}{dx^3}\pd{f^1}{u^2}\\&-
\frac 1{n} ({u^1}^2+{u^2}^2-1)
\frac {dn}{dx^3}(u^1\pd{f^1}{u^1}+u^2\pd{f^1}{u^2})=1,
\endaligned
$$
and the adjoint system is
$$\frac{dx^1}{u^1}=\frac{dx^2}{u^2}=\frac{dx^3}{1}=
-\frac{n\, du^1}{u^1({u^1}^2+{u^2}^2+1)\frac {dn}{dx^3}}=
-\frac{n\, du^2}{u^2({u^1}^2+{u^2}^2+1)\frac {dn}{dx^3}}=\frac {df^1}{1}.
$$

>From here we see that $f^1$ is $f^1=x^3+\varphi(x^1,x^2,u^1, u^2)$
where $\varphi$ is an arbitrary function of the first integrals of 
the adjoint system
$$\frac{dx^1}{u^1}=\frac{dx^2}{u^2}=\frac{dx^3}{1}=-\frac{n\, du^1}
{u^1({u^1}^2+{u^2}^2+1)\frac {dn}{dx^3}}=
-\frac{n\, du^2}{u^2({u^1}^2+{u^2}^2+1)\frac {dn}{dx^3}}.
$$

We see that $C_1=\frac {u^1}{u^2}$ is one of such first integrals and then 
$C_2=x^1-\frac {u^1}{u^2} x^2$ is a second one. The two last first 
integrals are to be obtained from 
$$
\frac{dx^2}{u^2}=\frac{dx^3}{1}=
-\frac{n \,du^2}{u^2((1+C_1^2){u^2}^2+1)\frac {dn}{dx^3}}, 
$$
and the solution depends on the concret choice for the function $n(x^3)$.
More specifically, we find 
$$C_3=n \exp\left[\int_0^{u^2}
\frac{  d\zeta }{\zeta((1+C_1^2){\zeta}^2+1)}\right],
$$ where we must replace after doing the computation the constant $C_1$ 
by the quotient $C_1=\frac {u^1}{u^2}$. The computation of the integral
is quite easy and leads to
$$C_3=n\sqrt{\frac{(1+C_1^2){u^2}^2}{(1+C_1^2){u^2}^2+1}}=n
\sqrt{\frac{{u^1}^2+{u^2}^2}
{{u^1}^2+{u^2}^2+1}}.$$
For every value of $C_3$ we
can express $u^{2}$ as a function of $n$, 
$$u^2=\frac{C_3}{\sqrt{(n^2-C_3^2)(1+C^2_1)}},
$$
and then the fourth first integral is given by
$$C_4=x^2-\int_0^{x^3} \frac{C_3}{\sqrt{(n^2(\zeta)-C_3^2)(1+C^2_1)}}\, 
d\zeta.
$$

In a similar way we can look for the function $F^2$ amongst the 
solutions of $KF^2=0$ and $\Delta F^2=1$. Solutions of $KF^2=0$ 
are the constants of motion and therefore $z_1=n^2v^1$, $z^2=n^2v^2$
which correspond to the momenta in the two first axes directions, where 
the medium index is constant, will give first integrals for $K$. Same
for $z_3=x_1-\frac {v_1}{v_2}x_2$, corresponding to the quotient of the
third component of the angular momentum (in Optics called {\sl skewness}
function or Petzval invariant) by the second component of
linear momentum
and  finally $z_4= n^2({v^1}^2+{v^2}^2+{v^3}^2)$, which is but the energy
of the Lagrangian $\Bbb L$. It is quite easy to see that
 $F^2=\log |\sqrt{z_4 }|$
is a solution of the system $KF^2=0$ and $\Delta F^2=1$.

According to this, we will do the following choice for the new coordinates:
$$\aligned y^1&=x^1-\frac {v^1}{v^2} \, x^2,\\  y^2&=x^2-\int _0^{x^3}
\frac {C_3}{\sqrt{(n^2(\zeta)-C_3^2)(1+C^2_1)}}\,d\zeta,\\  y^3&=x^3\\
y^4&=\frac {nv^1}{\sqrt{ {v^1}^2+{v^2}^2+{v^3}^2}},\\ 
y^5&=\frac {nv^2}{\sqrt{{v^1}^2+{v^2}^2+{v^3}^2}},\\ 
y^6&=\log \left[n\sqrt {{v^1}^2+{v^2}^2+{v^3}^2}\,\right],
\endaligned
\tag 5.6
$$
\vfil\eject 
namely, the inverse change is given by

$$\aligned
x^1&=y^1+\frac {v^1}{v^2}\left(y^1+\int _0^{y^3}
\frac {C_3}{\sqrt{(n^2(\zeta)-C_3^2)(1+C^2_1)}}\,d\zeta \right),\\
x^2&=y^2+\int _0^{y^3}
\frac {C_3}{\sqrt{(n^2(\zeta)-C_3^2)(1+C^2_1)}}\,d\zeta,\\
x^3&=y^3,\\
v^1&=\frac 1{n^2}\,e^{y^6}y^4,\\
v^2&=\frac 1{n^2}\,e^{y^6}y^5\\
v^3&=\frac 1{n^2}\,e^{y^6}\sqrt{n^2-{y^4}^2-{y^5}^2}
\endaligned
\tag 5.7
$$
and if we recall the expression of $\w_L$ written in the form
$$
\aligned
\w_L&=dx^1\wedge d\left(\frac {nv^1}{\sqrt {{v^1}^2+{v^2}^2+{v^3}^2}}\right)
+dx^2\wedge d\left(\frac {nv^2}{\sqrt {{v^1}^2+{v^2}^2+{v^3}^2}}\right)
\\&+\frac {n{v^3}^2}{\left({v^1}^2+{v^2}^2+{v^3}^2\right)^{3/2}}
\left[v^1d\left(
\frac {v^1}{v^3}\right)\wedge dx^3+
v^2d\left(
\frac {v^2}{v^3}\right)\wedge dx^3\right]
\endaligned
$$
we will get the new coordinate expression
$$\aligned
\w_L=&\left[dy^1+d\left(\frac {y^4}{y^5}y^2\right)+
d\left(\frac {y^4}{y^5}\int_0^{y^3}\frac {C_3}{\sqrt{(n^2(\zeta)-C_3^2)
(1+C^2_1)}}\,d\zeta\right)\right]\wedge dy^4\\+
&\left[dy^2+d\int_0^{y^3}\frac {C_3}{\sqrt{(n^2(\zeta)-C_3^2)
(1+C^2_1)}}\,d\zeta\right]\wedge dy^5\\+&
\frac 1{(n^2-{y^4}^2-{y^5}^2)^{1/2}}\left[y^4\,dy^4\wedge dy^3+y^5\,
dy^5\wedge dy^3\right],
\endaligned
$$
and therefore, when developing these expressions we will find that some 
terms cancel and only remain
$$\w_L=d\left(y^1+\frac {y^4}{y^5}\,y^2\right)\wedge dy^4
+dy^2\wedge dy^5,\tag 5.8
$$
which shows that 
$$\aligned \xi^1&=y^1+\frac {y^4}{y^5}y^2=x^1-\frac {v^1}{v^2}
\int_0^{x^3}\frac {C_3}{\sqrt{(n^2(\zeta)-C_3^2)
(1+C^2_1)}}\,d\zeta,\\ \xi^2&=x^2-\int_0^{x^3}\frac {C_3}
{\sqrt{(n^2(\zeta)-C_3^2)
(1+C^2_1)}}\,d\zeta
\endaligned\tag 5.9
$$
and the corresponding 
$$\eta^1=y^4=\frac {nv^1}{\sqrt {{v^1}^2+{v^2}^2+{v^3}^2}},\quad 
\eta^2=y^5=\frac {nv^2}{\sqrt {{v^1}^2+{v^2}^2+{v^3}^2}}\tag 5.10
$$
are Darboux coordinates for the symplectic form induced in the quotient
space.

Let us now consider the 
 particular but important case  case in which  the refractive index becomes
 constant out of  a region. If for $x^3>L$, the index $n$ is constant,
 the above mentioned Darboux coordinates $\xi^1$ and $\xi^2$ are
$$
\xi^1=x^1-\frac {v^1}{v^2} \frac {C_3}{\sqrt{(n^2-C_3^2)
(1+C^2_1)}},\quad \xi^2=x^2- \frac {C_3}{\sqrt{(n^2-C_3^2)
(1+C^2_1)}},
$$
up to a constant, and from  
$$C_1= \frac {v^1}{v^2},\quad  \frac {C_3}{\sqrt{(n^2-C_3^2)(1+C_1^2)}}=
\frac {v^2}{v^3},
$$
we see  that the Darboux coordinates become 
$$x^1-\frac {v^1}{v^3}x^3,\quad x^2-\frac {v^2}{v^3}x^3, \quad \frac {nv^1}
{\sqrt{{v^1}^2+{v^2}^2+{v^3}^2}},\quad  \frac {nv^2}
{\sqrt{{v^1}^2+{v^2}^2+{v^3}^2}},\tag 5.11
$$
in full agreement with (4.9). Therefore, for an optical system such that
the refractive index depends only on $x^3$ and, furthermore, 
 the region in which the index is not constant is bounded, we can choose
Darboux coordinates by fixing a $x^3$ outside this region and taking Darboux 
coordinates for the corresponding problem of constant index. This justify
the choice of coordinates as usually done for the  ingoing and outgoing light rays in the
corresponding constant index media, i.e. it shows the convenience of using flat screens
in far enough regions on the left and right respectively, and then this 
change of Darboux 
coordinates seems to be, from an active viewpoint, 
a canonical transformation.
    
Let us now consider a different and interesting  particular case in which 
there exists a symmetry axis and $n$  depends on the distance to 
this axis alone. In this case the
ray paths describe perfect optical instruments  \cite {\HolmKo}. We will use cylindrical coordinates $(r,\theta,z)$ and then 
the metric  has nonzero elements $g_{rr}=n^2,\,g_{\theta\theta}=r^2n^2$, 
and $g_{zz}=n^2$. The nonzero Christoffel symbols are
$$
\aligned
\G^r_{rr}&=\frac 1n \frac {dn}{dr}\\
\G^r_{\theta\theta}&=-\frac 1{2n^2}\frac d{dr}(r^2n^2)=-r-\frac {r^2}n
\frac {dn}{dr}\\
\G^r_{zz}&=-\frac 1n \frac {dn}{dr}\\
\G^\theta_{r\theta}&=\G^\theta_{\theta r}=\frac 1r+\frac 1n  \frac {dn}{dr}\\
\G^z_{rz}&=\G^z_{zr}=\frac 1n \frac {dn}{dr},
\endaligned
\tag 5.12
$$
and therefore $\G_{\Bbb L}$ is given by
$$\aligned
\G_{\Bbb L}&= v_r \pd {}r+ v_\theta \pd{}{\theta}+ v_z \pd {}z+\left[\frac 1n  
\frac {dn}{dr}\left(r^2 v_\theta ^2+ v_z^2- v_r ^2\right)+r v^2_\theta \right]
\pd{}{ v_r}\\&+
\left[-\frac 2n\frac {dn}{dr} v_\theta  v_r-2\frac{ v_r v_\theta}r
\right]\pd {}{ v_\theta}+
\left[-\frac 2n\frac {dn}{dr} v_\theta  v_r\right]\pd{}{ v_z}.
\endaligned
\tag 5.13
$$

Here we will take $K$ as being given by $\frac 1{v_r}\G_{\Bbb L}$. The search
for  coordinates adapted to the distribution $\Ker \w_L$ are
found as in the previous example. A particular choice is:
$$y^1=r^2\,\frac { v_\theta}{ v_z},\quad y^2=\frac {\sqrt{ v_r^2+r^2
v_\theta^2+ v_z^2}}
{n v_z}, \quad y^3=r,$$
and 
$$y^4=\theta-\int_0^r\frac {k\,dr}{r^2\sqrt{c^2n^2-\frac {k^2}{r^2}-1}},\,
y^5=z-\int_0^r\frac {dr}{\sqrt{c^2n^2-\frac {k^2}{r^2}-1}},\, y^6=\log n^2 v_z.
$$
where $c$ and $k$ are the constants of motion 
$$k=y^1=r^2
\frac { v_\theta}{ v_z}\ \text{ and }\ c=y^2=\frac {\sqrt{ v_r^2+r^2 v_\theta^2+
v_z^2}}
{n v_z}.$$

The presymplectic form $\w_L$,
$$\aligned
\w_L&=d\theta\wedge d\left(\frac {nr^2 v_\theta}{\sqrt{ v^2_r+r^2 v_\theta^2+ v_z^2}}
\right)+dr\wedge d\left(\frac {n v_r }{\sqrt{ v^2_r+r^2 v_\theta^2+ v_z^2}}
\right)\\&+dz\wedge d\left(\frac {n v_z}{\sqrt{ v^2_r+r^2 v_\theta^2+ v_z^2}}
\right),
\endaligned
$$
when written using the new coordinates will become after cancellation of
some terms
$$\w_L=dy^4\wedge d\left(\frac {nr^2 v_\theta}
{\sqrt{ v_r^2+r^2 v_\theta^2+ v_z^2}}\right)
+dy^5\wedge d\left(\frac {n v_z}{\sqrt{ v_r^2+r^2 v_\theta^2+ v_z^2}}\right),
\tag 5.14$$
from which we see that Darboux coordinates for the reduced symplectic
form in the quotient space are
$$\xi^1=y^4=\theta-\int_0^r\frac {k\,dr}{r^2\sqrt{c^2n^2-\frac {k^2}{r^2}-1}},\,
\xi^2=y^5=z-\int_0^r\frac {dr}{\sqrt{c^2n^2-\frac {k^2}{r^2}-1}},
\tag 5.15
$$ and the corresponding ones 
$$\eta^1=\frac {nr^2 v_\theta}
{\sqrt{ v_r^2+r^2 v_\theta^2+ v_z^2}},\quad \eta^2=\frac {n v_z}
{\sqrt{ v_r^2+r^2 v_\theta^2+ v_z^2}},
\tag 5.16
$$
which reduces to (4.11) when $n$ is constant.   

\head
6. Outlook
\endhead

The theory here developed   
  suggests the interest of the study of what happens 
for anisotropic media, because of the recent ineterest in the use of anisotropic
optical material. 
   The theory can be  reexamined  along similar lines, because 
the basic principle of the theory is still the celebrated Fermat's principle
of least time (or optical time if reflection is also allowed). 
 The only difference is that   when the medium is not isotropic
the  refractive index of the medium  (given by the quotient $n=\frac cv$), 
   may depend on the direction of 
the ray, and then $\Delta n=0$.  

Again we will use the analogous mechanical  problem 
 where the Lagrangian function 
is given by $L=n\, \sqrt {g(v,v)}$.
This Lagrangian    is still homogeneus 
of degree one and the corresponding energy vanish identically, and therefore
the Lagrangian is singular.  
 It is possible to show that, at least when the 
Lagrangian 
$\Bbb L=\frac 12 n^2 \,g(v,v)$ is regular,  
 the curves 
solution for  the  Lagrangian $\Bbb L$
are just the curve solution of the original 
problem,
even if the curves are reparametrized.  Even more, the space of light rays in this 
particular case is also a symplectic manifold. This problem and the search for 
appropriate Darboux coordinates will be examined in a forthcoming paper. 
 
\bigskip

\noindent{\bf Acknowledgements:} 
  Partial financial support from
DGICYT,  Research Grants PS93--0582, is
acknowledged.

\bigskip

{\bf References} 

\refno\Saletan.  
E.J. Saletan  and  A.H. Cromer,
{\subrayar{Theoretical Mechanics}} (J. Wiley, 1971)

\refno\AbM.
 R. Abraham   and J.E. Marsden,
\subrayar{Foundations of Mechanics}
(Reading, Ma: Benjamin, 1978) 
  
\refno\GS. 
V. Guillemin  and S. Sternberg, 
\subrayar{Symplectic Techniques in Physics}
(Cambridge U.P., London, 1984)
 
\refno\Bamberg. 
P. Bamberg  and S. Sternberg,
\subrayar{A course on Mathematics for Students of Physics}
(Cambridge U.P., London, 1988)
 
\refno\SW.
J.  S\'anchez Mondrag\'on and K.B. Wolf, 
\subrayar{Lie Methods in Optics},
Lect. Notes in Physics 250, 
(Springer,  1986)
 
\refno\Wolfuno. 
K.B. Wolf 
\subrayar{Lie Methods in Optics II},
 Lect. Notes in Physics 352, 
(Springer, 1989)
 
\refno\Dragtuno. 
A.J. Dragt  and  J.M. Finn,
``Lie series and invariant functions for analytical symplectic maps",
 J. Math. Phys. 
{\bf 17}, 2215--27 (1976)
 
\refno\Dragtdos.
 A.J. Dragt, 
``Lie algebraic theory of geometrical optics and optical aberrations",
J. Opt. Soc. Amer. 
{\bf 72}, 372--79 (1982)

\refno\Dragttres.
 A.J. Dragt,
``Lectures on nonlinear Orbit Dynamics" in
\subrayar{Physics of High Energy} \subrayar{ Particle Accelerators},
  A.I.P. Conferences Proceedings No. 87. 
(Amer. Inst. Phys., New York, 1982)
 
\refno\DForest. 
 A.J. Dragt   and E. Forest,
``Computation of nonlinear behavior of Hamiltonian systems using Lie
algebraic methods", J. Math. Phys. 
{\bf 24}, 2734--44 (1983)
 
\refno\Wolfdos. 
 K.B.  Wolf, 
``Symmetry in Lie optics",
  Ann. Phys.  {\bf  172}  1--25 (1986)
 
\refno\CL. 
O.  Casta\~nos, E. L\'opez Moreno  and K.B. Wolf, 
``Canonical transforms for paraxial wave optics", in
\subrayar{Lie Methods in Optics}, J. S\'anchez Mondrag\'on  and  K.B. Wolf,
 Lect. Notes in Physics 250 
(Springer, 1986)
 
\refno\DFW. 
   A.J. Dragt, E. Forest  and K.B. Wolf,
``Foundations of a Lie algebraic theory of geometrical optics",
 in \subrayar{Lie Methods in Optics}, J. S\'anchez Mondrag\'on  and K.B. 
 Wolf,
 Lect. Notes in Physics 250 (Springer, 1986)

\refno\Pepin. 
 J.F. Cari\~nena   and  C. L\'opez,  
``Symplectic structure on the set of geodesics of a Riemannian manifold",
Int. J. Mod. Phys. {\bf 6}, 431--44 (1991)

\refno\Pepindos.
 J.F. Cari\~nena,  
``Theory of Singular Lagrangians", 
   Forts. der Phys. {\bf 38},
  641--680  (1990)

\refno\Marsden.
 J.E. Marsden  and  A. Weinstein,
``Reduction of symplectic manifolds with symmetry",
Rep. Math. Phys.
{\bf 5}, 121--30  (1974)
 
\refno\HolmKo. 
 D.D. Holm   and G. Kovacic,
``Homoclinic Chaos for Ray Optics in 
a fiber",   Physica D
{\bf 51}, 177--188 (1991)

\refno\HolmWo.  
 D.D. Holm   and K.B. Wolf,
``Lie Poisson description of Hamiltonian ray optics", Physica D
{\bf 51}, 189--199
(1991)

\refno\HolmKov.       
  D.D. Holm    and G. Kovacic,
``Homoclinic Chaos from Ray Optics in 
a fiber: 200 Years after Lagrange", in \subrayar{Mechanics, Analysis and  
Geometry: 200
 Years} \subrayar{after Lagrange},
  Ed: M. Francaviglia (Elsevier, 1991)
 
\refno\Wolftres. 
 K.B. Wolf, ``Nonlinearity in Aberration Optic", in 
 \subrayar{Symmetries and Nonlinear} \subrayar{Phenomena},
 Eds:  D. Levi D. and P. Winternitz, (World Sci. Press, 1988)

\refno\CN.
J.F. Cari\~nena,  J. Fern\'andez-N\'u\~nez and 
 E. Mart\'{\i}nez,
 ``A geometric approach to Noether's Second  Theorem  in
time-dependent Lagrangian Mechanics", Lett. Math. Phys.
{\bf 23}, 51--63 (1991)

\refno\CP. 
 M. Crampin M. and  F.A.E. Pirani,
\subrayar{Applicable differential geometry}
(Cambridge U.P., London, 1986)  
 
} 
\enddocument